\documentclass[epj,referee]{svjour}
\usepackage{graphics}
\begin{document}

\title{Diversity of reproduction rate supports cooperation in the prisoner's dilemma game on complex networks}

\author{Attila Szolnoki\inst{1} \and Matja{\v z} Perc\inst{2} \and Gy{\"o}rgy Szab{\'o}\inst{1}}

\institute{Research Institute for Technical Physics and Materials Science, 
P.O. Box 49, H-1525 Budapest, Hungary  
\and Department of Physics, Faculty of Natural Sciences and Mathematics, University of Maribor, 
Koro{\v s}ka cesta 160, SI-2000 Maribor, Slovenia}

\date{Received: date / Revised version: date}

\authorrunning{A. Szolnoki, et al.}
\titlerunning{Diversity of reproduction supports cooperation in the prisoner's dilemma game...}

\abstract{
In human societies the probability of strategy adoption from a given person may be affected by the personal features. Now we investigate how an artificially imposed restricted ability to reproduce, overruling ones fitness, affects an evolutionary process. For this purpose we employ the evolutionary prisoner's dilemma game on different complex graphs. Reproduction restrictions can have a facilitative effect on the evolution of cooperation that sets in irrespective of particularities of the interaction network. Indeed, an appropriate fraction of less fertile individuals may lead to full supremacy of cooperators where otherwise defection would be widespread. By studying cooperation levels within the group of individuals having full reproduction capabilities, we reveal that the recent mechanism for the promotion of cooperation is conceptually similar to the one reported previously for scale-free networks. Our results suggest that the diversity in the reproduction capability, related to inherently different attitudes of individuals, can enforce the emergence of cooperative behavior among selfish competitors.
}

\maketitle

\section{Introduction}\label{sec:introduction}
Evolutionary game theory is a successful paradigm for studying interactions
among individuals as different as bacteria \cite{kerr_n02} and humans \cite{hofbauer_88}.
One branch of game theory considers the problem of cooperation as a particular example of such interactions. The conflict between the individual and common interests is frequently modeled by the so-called prisoner's dilemma game \cite{axelrod_84}. Originally the game consists of two players who have to decide simultaneously whether they want to cooperate or defect. Mutual cooperation yields the highest collective benefit shared equally between the players. However, a defector can have a higher individual payoff if the opponent decides to cooperate. Therefore both players decide to defect, whereby they end up with a lower payoff than if both would cooperate; hence the dilemma. This unfavorable result of classical game theory is, however, often at odds with reality \cite{sigmund_94}. Accordingly, several mechanisms, ranging from kin-selection to various forms of reciprocity \cite{nowak_s06} and other more sophisticated processes \cite{wang_arx07}, have been proposed to explain the emergence of cooperation. Particularly inspiring in the latter aspect, and still widely investigated, is also the spatial extension of the classical prisoner's dilemma game \cite{nowak_n92a,lindgren_pd94,schweitzer_acs02} as well as other games with different payoff rankings \cite{nowak_ijbc94}. Although the outcome of so-called games on grids depends somewhat on their numerical implementation \cite{frean_prsb94}, the cooperation-facilitating effect in the context of the prisoner's dilemma game is robust. 

The success of the spatial prisoner's dilemma game to sustain cooperation has made it a common starting point for further explorations of mechanisms that could facilitate cooperation even beyond the borders determined solely by the spatial extension. For example, it proved very successful to introduce a third strategy into the game. The so-called loners, or volunteers, induce a rock-scissors-paper-type cyclic dominance of the three strategies \cite{frean_prsb01} and are able to prevent, via oscillatory changes, the extinction of cooperators even by high temptations to defect \cite{hauert_s02,szabo_prl02}. Noteworthy, the loners also promote cooperation in the absence of spatial interactions. Moreover, the impact of variable degrees of investment in the prisoner's dilemma game has also been studied, as was the suitable walk of agents on the grid \cite{vainstein_jtb07}, as well as fine-tuning of noise and uncertainties by strategy adoptions \cite{szabo_pre05,perc_njp06a,perc_pre07a}.

To exceed the simplest lattice graphs, more specific topologies of networks defining the interactions among individuals were also studied \cite{abramson_pre01,ohtsuki_prsb06}, which has received substantial attention (for a review see \cite{szabo_pr07}). More specifically, the celebrated scale-free graph has been recognized as an extremely potent promoter of cooperative behavior in the prisoner's dilemma as well as in the snowdrift game \cite{santos_prl05,santos_prsb06}, and this promotion of cooperation has been found robust on several factors \cite{poncela_njp07}. However, one may argue that the many links of a hub involve not just a higher payoff but a higher cost as well. Therefore, the use of normalized payoffs may represent a more realistic approach \cite{santos_jeb06}. Indeed, the introduction of participation costs eradicates the ability of scale-free networks to promote cooperation \cite{masuda_prsb07}, yielding similar levels of cooperative behavior as regular grids introduced initially by Nowak and May \cite{nowak_n92a}. Very recently, Pacheco {\it et al.} have reported that a suitable dynamical linking helps to maintain cooperative behavior \cite{pacheco_prl06,pacheco_jtb06}, whereas on the other hand, Ohtsuki {\it et al.} have shown that the separation of the interaction and strategy adoption graphs completely disables the survival of cooperators if the overlap between the two graph is zero \cite{ohtsuki_prl07,ohtsuki_jtb07}. Inhomogeneities in the strategy adoption probabilities can also enhance the frequency of cooperators \cite{wu_cpl06,wu_pre06}, particularly if the strategy adoption is favored from some distinguished players \cite{szolnoki_epl07}. It is worth mentioning that the introduction of other inhomogeneities in the personality (e.g., stochastic payoffs \cite{traulsen_jtb07}, different aspiration levels for the win-stay-lose-sift strategies \cite{posch_prslb99}, or even individual acceptance levels for the evolutionary Ultimatum games \cite{sanchez_jtb05}) can also support the altruistic behavior under certain conditions.  

In this paper, we extend the above investigations by introducing the diversity of reproduction of individuals on two types of complex networks; namely on regular small-world graphs and on highly irregular scale-free networks. To study the impact of the diversity of reproduction on the stationary level of cooperation explicitly, we focus on normalized payoffs by the prisoner's dilemma game. We find that the differences of reproduction facilitate cooperative behavior irrespective of the interaction network and particularities concerning payoff accumulation, and moreover, may lead to domination of cooperation where otherwise defection would reign. Remarkably, although normalized payoffs can eliminate the advantage of scale-free topology, we observe that reproduction restrictions of players that are inversely proportional to their connectivity restore the cooperative trait across the whole parameter range of the temptation to defect. By studying cooperation levels within the group of individuals having full reproduction capabilities, and comparing those to the overall fraction of cooperators, we are able to draw strong parallels between the presented mechanism for the promotion of cooperation and the one reported previously for scale-free networks. Indeed, our results imply that several recently introduced mechanisms for the promotion of cooperation within the prisoner's dilemma game are routed in diversities of participating players, which may emerge intrinsically due to an inhomogeneous interaction network \cite{santos_prsb06}, or can be introduced extrinsically via social diversity \cite{perc_arx07} or reproduction restrictions.

The remainder of this paper is structured as follows. Section~\ref{sec:PD} is devoted to the description of particularities of the evolutionary prisoner's dilemma game and reproduction restrictions on small-world and scale-free networks, while Section~\ref{sec:result} features the results. In the last Section we outline potential implications of our findings.

\section{Evolutionary prisoner's dilemma game}
\label{sec:PD} 

We consider an evolutionary two-strategy prisoner's dilemma game with players located on vertices of either regular small-world graphs or irregular scale-free networks. Via the analogy with the creation of the Watts-Strogatz structure, the former graph is generated from a regular two-dimensional grid by randomly rewiring a certain fraction $Q$ of nearest-neighbor links whereby preserving the initial connectivity $z_x=4$ of each player $x$ \cite{szabo_jpa04}. Evidently, for $Q=0$ this structure is a square lattice, whereas the limit $Q \to 1$ yields a regular random graph. The scale-free network is generated via the celebrated mechanism of preferential attachment growth \cite{barabasi_s99} yielding a power-law distribution of $z_x$ but still having average connectivity $z=4$. Initially, each player $x$ is designated as a cooperator ($C$) or defector ($D$) with equal probability. Moreover, amongst all $N$ players, and irrespective of their initial strategies, a fraction $\nu$ of players is chosen randomly and designated as having a restricted ability to transfer their strategy \cite{szolnoki_epl07}. The parameter $\nu$ is crucial in the present work since it determines the fraction of players having restricted reproduction capabilities, and accordingly, will be in the focus of simulation results presented in the following Section. Importantly, the reproduction ability of each player is set only once at the beginning of each simulation and remains unchanged during the evolutionary process. Next, a player $y$ can reproduce its strategy $s_y$ on one of its randomly chosen neighbors $x$ (throughout this work "neighbors of $y$" refers to those which are directly connected with $y$) in accordance with the probability
\begin{equation}
W(s_x \leftarrow s_y)=W_y\,\frac{1}{1+\exp[(P_x-P_y)/K]}\,\,,
\label{eq:prob}
\end{equation}
where $W_y=w<1$ if player $y$ has a restricted ability to transfer its strategy, and $W_y=1$ otherwise. Note that the choice of $w=0$ in the former case would mean that $\nu N$ players amongst all $N$ are completely unable to reproduce their own strategy, which would lead to frozen states and stop the evolutionary process for large enough $\nu$. It is also easy to see that the stationary state at $\nu=0$ agrees with the state at $\nu=1$ but the relaxation is slower in the latter case. Moreover, $K$ characterizes the uncertainty related to the reproduction process, also serving to avoid trapped conditions and warranting smooth transitions towards stationary states. Payoffs $P_x$ and $P_y$ of both players are calculated in accordance with the standard prisoner's dilemma scheme \cite{nowak_n92a} having temptation $b$, reward $1$, and both punishment as well as the suckers payoff $0$, where $1 < b \leq 2$ to ensure a proper payoff ranking. More precisely, both players $x$ and $y$ play one round of the prisoner's dilemma game with all their neighbors, respectively. Their accumulated payoffs resulting from $z_x$ and $z_y$ interactions are stored in $p_x$ and $p_y$. As mentioned above, these payoffs are normalized with the number of interactions from which they were obtained, hence yielding $P_x=p_x/z_x$ and $P_y=p_y/z_y$. Using normalized payoffs we can separate effects described earlier by Santos {\it et al.} \cite{santos_prl05,santos_prsb06} from those appearing due to the presently introduced  reproduction restrictions.  

The elementary steps of the game on the two considered types of complex networks are as follows. An arbitrarily chosen player $x$ acquires its payoff $P_x$ by playing the game with all its neighbors. One randomly chosen neighbor of player $x$; we denote it by $y$, also acquires its payoff $P_y$ by playing the game with all its neighbors. Finally, the reproduction of player $y$ to the site of player $x$ is attempted according to Eq.~(\ref{eq:prob}). A full Monte Carlo step (MCS) consists of executing the above-described elementary steps $N$ times. 
Simulations of the evolutionary process via the Monte Carlo algorithm were performed for populations comprising $N=10^5 - 10^6$ players, and characteristic quantities, such as the stationary frequencies of cooperators $\rho_c$ and defectors $\rho_d$, were averaged over a sampling period ($t_s=10^4 - 10^6$ MCS) after a sufficiently long transient time $t_r \approx t_s$.

\section{Results}
\label{sec:result} 

\begin{figure}
\resizebox{0.95\columnwidth}{!}{\includegraphics{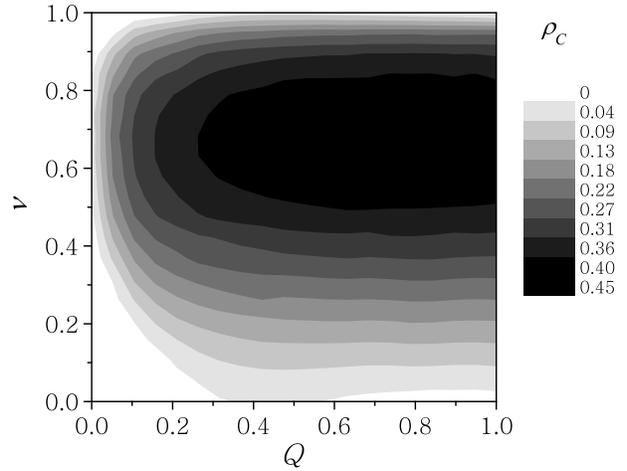}}
\caption{Gray scale coded fraction of cooperators $\rho_c$ in dependence on $Q$ and $\nu$, obtained for $K=0.08$ and $b=1.14$. The gray scale is linear, white depicting 0.0 and black 0.45 values of $\rho_c$.}
\label{fig:phd}
\end{figure}

The above described mechanism as a type of inhomogeneous teaching activity in a social context has proved to promote cooperation on lattices \cite{szolnoki_epl07}. Here we extend this study to explore the possible role of the topological features of complex graphs. To have an overview of the impact of restricted reproducibility on a regular small-world graph we calculated $\rho_c$ systematically for different values of $Q$ and $\nu$ while the values of $K$ and $b$ were held fixed. A typical contour plot of cooperation is plotted in Fig.~\ref{fig:phd}, where $K=0.08$ and $b=1.14$ were used. Within a wide region of $\nu$ the level of cooperation ($\rho_c$) is enhanced significantly and this improvement increases monotonously with the randomness of the interaction topology via $Q$. In particular, the fraction of cooperation can be enhanced from $\rho_c=0$ (for $\nu=0$ and $Q=0$) to $\rho_c \approx 0.45$ if $\nu=0.6$ and $Q > 0.5$. The cooperation level saturates if the rewiring probability exceeds the value $Q \approx 0.5$. Notice that the optimal value of $\nu$ is practically independent of $Q$ ($\nu_{opt} \approx 0.6$).

\begin{figure}
\resizebox{0.95\columnwidth}{!}{\includegraphics{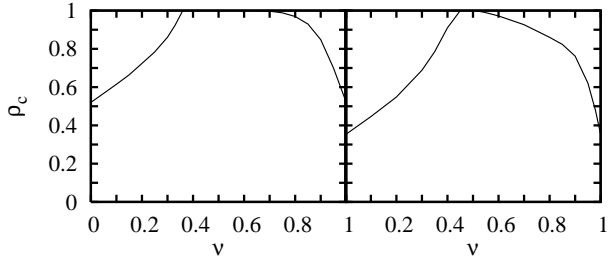}}
\caption{Fraction $\rho_c$ of cooperators when varying $\nu$: (left) on the regular small-world graph generated with $Q=0.1$ for $K=0.4$ and $b=1.02$; (right) on the scale-free network for $K=0.1$ and $b=1.04$.}
\label{fig:sw}
\end{figure}

The Monte Carlo simulations show that reproduction restrictions expand and shift the region of parameters for which the cooperator and defector strategies coexist. We emphasize that defectors can die out completely ($\rho_c=1$) below a threshold value of $b$, whereby the latter depends on $Q$, $K$, $\nu$ and $w$. Figure~\ref{fig:sw}(a) demonstrates the effect of $\nu$ on the fraction of cooperators on a regular small-world structure for $Q=0.1$. Notably, for sufficiently low values of $w$ there exists a central region of $\nu$ for which $\rho_c=1$. For larger values of $w$ this region disappears and one can only observe a peak in the contour of $\rho_c(\nu)$.

To extend the studied class of connection topologies we have also analyzed the effect of reproduction restrictions on the maintenance of cooperation for the strongly degree-inhomogeneous Barab{\'a}si-Albert scale-free network \cite{barabasi_s99}. Figure~\ref{fig:sw}(b) shows that the fraction of cooperators remains fairly low for $b=1.04$ if $\nu=0$. However, as soon as $\nu$ is increased the fraction of cooperators rises quickly, eventually reaching $\rho_c=1$ at $\nu \approx 0.5$. Noteworthy, the optimal fraction of players suffering under reproduction restrictions on the scale-free network is comparable to the one identified for regular small-world networks. The similarity of behavior on these networks is related to the usage of normalized payoffs that suppress the otherwise important role of hubs by the scale-free connectivity structure \cite{santos_jeb06,wu_pa07}.

\begin{figure}
\resizebox{0.95\columnwidth}{!}{\includegraphics{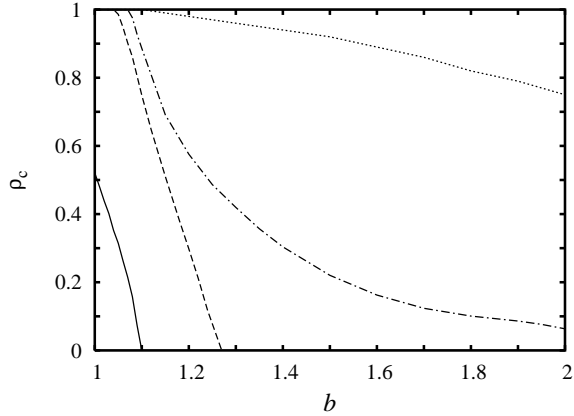}}
\caption{Fraction of cooperators $\rho_c$ as a function of $b$ when $\nu=0$ (solid line), 
$\nu=0.5$ (dashed line), and when the reproduction restriction is proportional to the connectivity of each player (dashed-dotted line) if $K=0.1$ The highest $\rho_c$ is achieved for the model suggested by \cite{santos_prl05} (dotted line).}
\label{fig:compare}
\end{figure}

Since an appropriately pronounced limitation of reproduction capabilities recovers the ability of scale-free networks to boost cooperation to dominance even if normalized payoffs are used in Eq.~(\ref{eq:prob}), it seems reasonable to investigate this phenomenon more precisely in dependence on $b$, and moreover, to test if there exist optimal ways of how reproduction restrictions can be introduced. Accordingly, the results presented in Fig.~\ref{fig:compare} compare $\rho_c$ as a function of $b$ for four different cases of reproduction restrictions. In the absence of reproduction restrictions ($\nu=0$) the scale-free network alone is unable to sustain cooperative behavior past $b=1.09$ since normalized payoffs are used for the evaluation of fitness. On the other hand, the region of coexisting $D$ and $C$ strategies is shifted towards significantly higher values of $b$; in particular, cooperators can survive up to $b=1.26$ if $\nu \approx 0.5$. 

Utilizing that scale-free networks have a power-law connectivity distribution, it is reasonable to investigate what happens if the reproduction capability of players is proportional to their degree $z_y$. For this purpose we have studied a system with site-dependent $w$, i.e. $w_y=z_y/z_{max}$ where $z_{max}$ is the maximal number of neighbors within the employed scale-free network for $\nu = 1$. These types of dynamical rules describe the situation when strategy adoption along each connection (in both direction) is allowed with the same probability. The dashed-dotted line in Fig.~\ref{fig:compare} illustrates that in this case cooperators survive throughout the whole range of $b$. The advance of players with large neighborhoods can be enhanced further either by comparing total payoffs \cite{santos_prl05} or by artificial preferences as suggested previously by Ren {\it et al.} \cite{ren_arx06}. To demonstrate the additional enhancement the dotted line in Fig.~\ref{fig:compare} shows the values of $\rho_c$ on scale-free networks when the players compare their absolute payoffs, as implemented in \cite{santos_prl05}. From the qualitative similarity between the last two cases one can suspect that similar mechanisms may underlie the enhancement of cooperative behavior for these systems. In order to clarify this assumption, we have studied the level of cooperation within the group of individuals having high reproduction capabilities, denoting this as $\rho_r$, that can be compared to the overall fraction of cooperators $\rho_c$. Evidently, the players with $w=1$ belong to the mentioned group in the case of randomly distributed players with two possible values of reproduction capabilities. However, when the reproduction capability is inversely proportional to the degree of a node several different values of $w$ are possible. Here we divide the nodes into three categories in such a way that each interval of the degree becomes equally large on the logarithmic scale. We consider players belonging to the "high reproduction capability class" if they are members of the highest connectivity group. 

\begin{figure}
\resizebox{0.95\columnwidth}{!}{\includegraphics{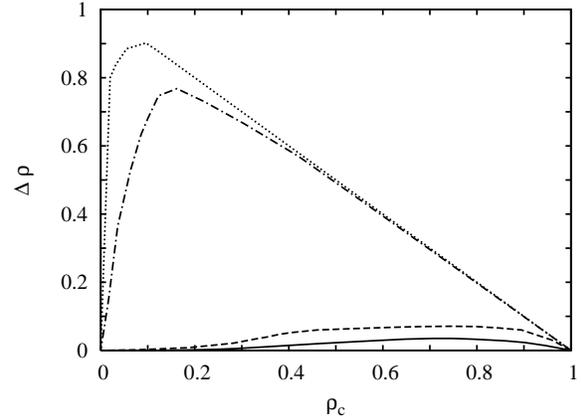}}
\caption{Excess fraction of cooperators $\Delta\rho$ within the group of individuals having high reproduction capability (or connectivity) versus the overall fraction $\rho_c$. The four curves (from bottom to top) illustrate the results obtained on the scale-free network with randomly distributed two-value reproduction capability (solid line), small-world network ($Q=0.9$) with randomly distributed two-value reproduction capability (dashed line), scale-free network with reproduction rate proportional to the connectivity of each player (dashed-dotted line), and scale-free network with absolute payoffs as in \cite{santos_prl05} (dotted line). The simulation were performed for $\nu=0.5$ and $K=0.1$.}
\label{fig:mutual}
\end{figure}

Results in Fig.~\ref{fig:mutual} illustrate the difference $\Delta\rho=\rho_r-\rho_c$ as a function of $\rho_c$ for the small-world graph and scale-free network at $\nu=0.5$, as well as for the case when reproduction restrictions are proportional to the connectivity of each individual (for comparative purposes results obtained with absolute payoffs and in the absence of reproduction restrictions on the scale-free network are also shown). In the latter case, and similarly to the inversely proportional restriction situation, players of the studied group have a connectivity belonging to the top third of the whole interval on a logarithmic scale.

The most important message of Fig.~\ref{fig:mutual} is that $\Delta \rho >0$ in the whole region of parameters. In other words, the cooperation is preferred on the sites from where the actual strategy can spread away faster independently of the mechanism yielding this process. Evidently, for pure cooperation ($\rho_c=1$) or defection ($\rho_c=0$) $\Delta\rho=0$. In the close vicinity of the boundaries the enhancement vanishes linearly, i.e., $\Delta  \rho \simeq \rho_c$ if $\rho_c << 1$ and  $\Delta  \rho \simeq (1 - \rho_c$) if $1-\rho_c << 1$. Between these two limits, however, values of $\Delta\rho$ indicate the enhancement of cooperation level within the group of individuals having high reproduction capabilities. 

The above results indicate that defection cannot survive long on network sites having high strategy reproduction capability. This phenomenon is analogous to the one reported by Santos {\it et al.} on the scale-free graph \cite{santos_prsb06}. In the latter case the total payoff difference determines the probability of strategy adoption that favors the sites with a high degree. Consequently, within the large neighborhood of a defector the fraction of defectors increases with time, and this process yields a monotonously decreasing total income for the focal defector. Sooner or later this focal defector will adopt the strategy from another focal cooperator whose imitation is beneficial for it, thereby, the focal cooperators become the players to be followed by others. In the present models a similar process takes place due to the introduction of inhomogeneous reproduction capabilities as an additional feature of players. Finally, we mention that the recently introduced promotion of cooperation via social diversity in the prisoner's dilemma game \cite{perc_arx07} also relies on a mechanism with the same properties as described above, hence implying it is widely applicable and may serve to identify new ways of avoiding widespread defection.

The above-described mechanism for the promotion of cooperation assumes that links between players having a large probability to affect their neighbor's strategy are rare \cite{santos_prsb06,rong_pre07}. Most notable enhancements of $\rho_c$ are warranted by the scale-free topology, which first, provides extremely high inhomogeneities in the strategy adoption probabilities, and second, yields practically unidirectional strategy adoptions between players having large and small connectivity. We have to emphasize, however, that on regular graphs the randomly distributed influential players (players with full reproduction ability) are connected to each other with an adequate probability only if their density is appropriately adjusted, i.e. neither too large nor too small. Despite this necessary condition, however, the simulations have indicated some increase in $\rho_c$ even if their density was very low or high, respectively.

\section{Summary}

We have studied the effect of inhomogeneous reproduction capabilities on the evolution of cooperation for the multi-agent evolutionary prisoner's dilemma game if the connectivity structure is described by different complex graphs, such as regular small-world graphs and strongly irregular scale-free networks. Our results suggest that the introduction of inhomogeneous reproduction capabilities, representing many realistic situations in human societies and animal communities, is a powerful and robust promoter of cooperative behavior that works irrespective of the complexity of the interaction network and other details concerning payoff accumulation and determination of fitness. Diversity in reproduction capability is a particularly potent promoter of cooperation if the connectivity structure ensures that many co-players follow the strategy of rarely linked but potent players as it happens on the scale-free graphs studied by Santos {\it et al.} \cite{santos_prl05,santos_prsb06}.

\begin{acknowledgement}
This work was supported by the Hungarian National Research Fund (T-47003) and the Slovenian Research Agency (grant Z1-9629). A. S. thanks Zsuzsa Danku for useful discussions.
\end{acknowledgement}

\end{document}